\documentstyle[preprint,aps,psfig]{revtex}
\newcommand{\beq}{\begin{equation}}
\newcommand{\eeq}{\end{equation}}
\newcommand{\beqa}{\begin{eqnarray}}
\newcommand{\eeqa}{\end{eqnarray}}
\newcommand{\ba}{\begin{array}}
\newcommand{\ea}{\end{array}}

\begin{document}

\draft



\widetext 

\title{Solitary-waves of the Nonpolynomial Schrodinger Equation: \\
Bright Solitons in Bose-Einstein Condensates} 
\author{Luca Salasnich} 
\address{
Dipartimento di Fisica, Universit\`a di Milano, \\
Istituto Nazionale per la Fisica della Materia, 
Unit\`a di Milano \\
Via Celoria 16, 20133 Milano, Italy\\
E-Mail: salasnich@mi.infm.it} 

\maketitle 

\begin{abstract} 
Recently we have derived an effective one-dimensional 
nonpolynomial Schr\"odinger equation (NPSE) 
[Phys. Rev. A {\bf 65}, 043614 (2002)] 
that accurately describes atomic Bose-Einstein 
condensates under transverse harmonic confinement. 
In this paper we analyze the stability of the 
solitary-wave solutions of NPSE by means of the 
Vakhitov-Kolokolov criterion. Stable self-focusing solutions 
of NPSE are precisely the 3D bright solitons experimentally 
found in Bose-Einstein condensates 
[Nature {\bf 417}, 150 (2002); Science {\bf 296}, 1290 (2002)]. 
These self-focusing solutions generalize the ones 
obtained with the standard nonlinear Schr\"odinger equation 
(NLSE) and take into account the formation of instabilities 
in the transverse direction. We prove that 
neither cubic nor cubic-quintic approximations of NPSE are able to 
reproduce the unstable branch of the solitary-wave solutions. 
Moreover, we show that the analitically found critical point 
of NPSE is in remarkable good agreement with recent numerical 
computations of 3D GPE. 
We also discuss the formation of multiple solitons 
by a sudden change of the sign of the nonlinear strength. 
This formation can be interpreted 
in terms of the modulational instability of the  
time-dependent macroscopic wave function of the Bose condensate. 
In particular, we derive an accurate analytical formula 
for the number of multiple bright solitons. 
\end{abstract} 

\vskip 4.cm 

\pacs{Invited paper to appear in \\
'Progress in Mathematical Physics Research' \\
(Nova Science Publishers, New York, 2004). }


\newpage 

\narrowtext

\section{Introduction} 

A nonlinear Schr\"odinger equation with nonpolynomial 
nonlinearity, the so-called 
nonpolynomial Schr\"odinger equation (NPSE), 
has been recently introduced and studied [1-3]. 
This equation has been derived in the context of Bose-Einstein 
condensation [4,5] from the 3D Gross-Pitaevskii equation [6,7] 
as a reliable effective 1D equation which describes the axial 
wavefunction $\psi(x,t)$ of a cigar-shaped condensate confined 
in the transverse direction $(y,z)$ by a harmonic 
trapping potential. 
NPSE has been successfully used to numerically investigate [8] the 
dynamics of Bose condensate solitary-waves experimentally observed 
by two groups [9,10]. 
In this letter we rigorously determine the linear stability domain 
of the solitary-wave solutions of NPSE by using the Vakhitov-Kolokolov 
criterion [11,12]. In particular, we find that NPSE admits both stable 
and unstable stationary solitary-wave solutions. 

\section{Generalized Schr\"odinger Equations}

For the generalized nonlinear Schr\"odinger equation (GNLSE), 
given by 
\beq 
\left[ i {\partial \over \partial t} + {1\over 2} 
{\partial^2 \over \partial x^2} + 
F(|\psi|^2) \right] \psi = 0 \; , 
\eeq
where $F(|\psi |^2)$ is the nonlinear term,  
the linear stability of solitary-waves is determined by the 
Vakhitov-Kolokolov criterion [11-13]. With the standard position 
\beq 
\psi(x,t)=\phi(x) \; e^{i\omega t} \; , 
\eeq 
where $\omega$ is the frequency of the fase and $\phi(x)$ is 
a real scalar field, from GNLSE one obtains the stationary second-order 
differential equation 
\beq 
\left[{1\over 2} {d^2\over dx^2} + F(\phi^2 ) \right] 
\phi = \omega \; \phi \; . 
\eeq
The first derivative $N_s'(\omega )$ of the number 
of particles invariant 
\beq 
N_s(\omega ) = \int dx \; \phi^2(x) \; , 
\eeq 
calculated for the soliton solution $\phi(x)$, allows one 
to predict the parameter 
region in $\omega$ where the soliton amplitude can grow or decay 
exponentially with a nonzero growth rate. The solitons are stable 
if $N_s'(\omega ) > 0$ and unstable if $N_s'(\omega ) < 0$ [13]. 

\section{Nonpolynomial Schr\"odinger equation}

In the case of the self-focusing nonpolynomial Schr\"odiger 
equation (NPSE) [1-3] the nonlinearity of Eq. (1) and Eq. (3) 
is given by 
\beq 
F(\phi^2) = {g\phi^2\over \sqrt{1-g\phi^2} } 
- {1\over 2} 
\left( {1\over \sqrt{1-g \phi^2} } 
+ \sqrt{1-g \phi^2} \right) \; , 
\eeq
where $g$ is the strength of the attractive interaction between 
particles in the Bose-Einstein condensate. In particular, 
the parameter $g$ is related to the s-wave scattering length $a_s$ 
and to the characteristic length $a_{\bot}$ of the transverse 
harmonic confinement by $g=2|a_s|/a_{\bot}$ [1-3,8]. 
\par 
Given a GNLSE like Eq. (3) it is quite natural to try 
a power expansion of $F(\phi^2)$. 
In the case of NPSE one finds 
\beq 
F(\phi^2)=-1 + g\phi^2 + {3\over 8} g^2 \phi^4 + ... \; .   
\eeq 
The lowest order in $g\phi^2$ gives the familiar 
one-dimensional self-focusing cubic NLSE. 
By including the next term of the power expansion one has 
a cubic-quintic NLSE, which has been 
investigated in several papers (see for instance [13]). 
But, as we shall show, the cubic-quintic 
approximation of NPSE is not able to correctly describe 
the stability of NPSE solitary waves. 
\par 
The exact stationary solitary-wave solutions of NPSE 
can be obtained in the following way. 
A simple constant of motion of the stationary NPSE is 
\beq 
E={1\over 2}\left({d\phi\over dx} \right)^2 + V(\phi ) \; ,    
\eeq 
where 
\beq 
V(\phi ) = - \omega \; \phi^2 -\phi^2\sqrt{1-g\phi^2} 
\eeq
is the effective potential energy of the system. 
In Figure 1 we plot the 
potential energy $V(\phi)$ of the full NPSE and 
its cubic and cubic-quintic approximations. 
As shown by Figure 1, for $\omega<0$ 
$V(\phi)$ is a double-well potential and the two 
approximations are quite close to 
the exact curve, apart near $\phi = \pm 1/\sqrt{g}$, 
the end-points of the NPSE which are related to the the singularity 
of the term $(1-g\phi^2)^{1/2}$. 
By imposing the boundary condition $\phi\to 0$ for 
$x \to \infty$ one finds $E=0$ and so 
\beq
{d\phi\over dx} = 
\sqrt{ 2 \phi^2 \sqrt{1- g\phi^2 } + 2 \omega \phi^2 } \; . 
\eeq  
The previous formula gives the integral equation 
\beq 
\int dx = 
\int d\phi {1\over 
\sqrt{ 2 \phi^2 \sqrt{1- g\phi^2 } + 2 \omega \phi^2 } } \; , 
\eeq 
from which one obtains the solitary-wave solution $\phi(x)$ 
written in implicit form (see also [8]) 
\beq 
\sqrt{2} x = {1 \over \sqrt{1+\omega}} \; 
arcth\left[ 
\sqrt{ \sqrt{1-g\phi^2}+\omega \over 1+\omega } \right] 
- {1 \over \sqrt{1-\omega}} \; 
arctg\left[ 
\sqrt{ \sqrt{1-g\phi^2}+\omega \over 1-\omega } 
\right] \; .   
\eeq  

\section{Vakhitov-Kolokolov stability criterion} 

The number of particles $N_s(\omega)$ of Eq. (4) can be calculated 
by observing that 
\beq 
dx =  d\phi \left( {d\phi\over dx} \right)^{-1} \; , 
\eeq
from which one gets 
\beq 
N_s(\omega ) = 
\int d\phi {\phi \over \sqrt{ 2 \sqrt{1- g\phi^2 } + 
2 \omega } } \; . 
\eeq 
This integral is analytically solvable and one obtains 
\beq 
N_s(\omega ) = {2\sqrt{2} \over 3 g} 
\left( 1 - 2 \omega \right) 
\sqrt{1+\omega} \; .  
\eeq 
In Figure 2 we plot the function $N_s(\omega )$ for some 
values of the nonlinear strength $g$. 
The Vakhitov-Kolokolov criterion [11,12] 
is based on the study of the sign of the first derivative 
\beq 
N_s'(\omega ) = -{\sqrt{2} \over g} 
{(1+2\omega ) \over \sqrt{1+\omega} }  
\eeq 
of the number of particle $N_s(\omega)$. One easly finds that 
the solitary-wave $\phi(x)$ given by Eq. (11) is linearly stable 
for $-1<\omega < -1/2$ and unstable for $-1/2<\omega <1/2$. 
Moreover $N_s(\omega)$ has its maximum value $N_s^{max}=4/(3g)$ 
at the critical frequency $\omega = -1/2$. 
\par 
In Figure 2 we also plot $N_s(\omega )$ obtained from 
the cubic NLSE and the cubic-quintic NLSE. 
In the case of cubic NLSE one finds 
\beq 
N_s(\omega ) = {2\sqrt{2}\over g} \sqrt{1+\omega} \; , 
\eeq 
from which one immediately obtains 
that the solitary-wave solutions are always stable. 
In the case of the cubic-quintic NLSE the 
number of particles is instead given by 
\beq 
N_s(\omega ) = {2\over g}\left( {\pi \over 2} - arctg\left[
{\sqrt{2} \over 2 \sqrt{1+\omega} } \right] \right) \; . 
\eeq 
Remarkably also in this case the 
Vakhitov-Kolokolov stability criterion tells 
us that there are no unstable solitary-waves. 
Thus the quintic term in the power expansion 
of NPSE is not sufficient to determine the stability 
domain of NPSE. 
\par 
As previously stated, in the context of Bose-Einstein condensation [1-3] 
the nonlinear strength $g$ is given by $g=2|a_s|/a_{\bot}$. 
By using this expression one finds that the maximum value 
of $N_s(\omega)$ for the full NPSE is given by 
\beq 
N_s^{max} = {2\over 3} {a_{\bot}\over |a_s|} \; . 
\eeq 
This value is very close to the critical number $N_c$ of attractive 
Bose-condensed atoms under transverse harmonic confinement found in [14] 
with a full numerical solution of the 3D GPE: 
\beq 
N_c = 0.676 {a_{\bot}\over |a_s|} \; . 
\eeq
Beyond $N_c$ there is the so-called 
collapse of the Bose-Einstein condensate. 
The collapse is due to the fact that the solitary-wave solution 
becomes linearly unstable, i.e. $N_s(\omega)<0$ in our NPSE. 
Note that this collapse has nothing to do with the singularity 
of the term $(1-g\phi^2)^{1/2}$ that is present in NPSE: 
in fact, also at the collapse the condition $g\phi^2 <1$ is satisfied. 

\section{Modulational instability} 

In a recent experiment [9] it has been reported the formation 
of bright solitons in a Bose-Einstein 
condensate of $^7$Li atoms induced by a sudden change in the sign 
of the scattering length from positive to negative.  
The formation of these solitons can be explained 
as due to the modulational instability 
of the time-dependent wave function of the Bose condensate, 
driven by imaginary Bogoliubov excitations [15]. 
\par 
Let us consider a Bose condensate with 
a repulsive inter-atomic interaction ($a_s>0$). 
The condensate is described by Eq. (1) and Eq. (5), 
setting $g'=-g$ instead of $g$ in Eq. (5). 
Under box axial confinement, the stationary homogeneous 
wave function is given by 
\beq 
\phi(x) = \sqrt{N\over L}  \; 
\eeq 
for $-L/2<x<L/2$ and zero elsewhere. $L$ is the size 
of the confining box and $N$ the total number of bosons. 
The Bogoliubov elementary excitations $\epsilon_k$  
of the static Bose condensate $\phi(x)$ 
are found by looking for solutions of the form 
\beq 
\psi(x,t) = e^{i\omega t}
\left[ \phi(x) + u_k(x) e^{-i\epsilon_k t}
+ v_k^*(x) e^{i\epsilon_k t}
\right] , 
\eeq 
and keeping terms linear in the complex functions 
$u(x)$ and $v(x)$ (linear-stability analysis). 
In the quasi-1D limit (defocusing NLSE) one finds 
\beq 
\epsilon_k = \sqrt{ {k^2\over 2} 
\left({k^2\over 2 } 
+ 2 g' n \right) }  , 
\eeq 
where $\omega = - g'n$, $n=N/L$ is the axial density 
and $g' = 2 a_s/a_{\bot}$. 
\par 
By suddenly changing the scattering length $a_s$ 
to a negative value, 
the excitations frequencies corresponding to 
\beq 
k<k_c = \sqrt{16 \pi |g'| n}
\eeq
become imaginary and, as a result, small perturbations grow 
exponentially in time. This penomenon is known as 
{\it modulational instability} [16,17]. 
It is easy to find that the maximum rate of growth is at 
$k_0=k_c/\sqrt{2}$. The wavelength of this mode is 
$\lambda_0 = {2\pi / k_0}$ and the ratio $L/\lambda_0$ gives 
an estimate of the number $N_{BS}$ of bright solitons which 
are generated: 
\beq 
N_{BS} = {\sqrt{N|a_s|L}\over \pi a_{\bot} } . 
\eeq 
The predicted number $N_{BS}$ of bright solitons 
is in very good agreement with the numerical results 
of 3D GPE [18] and in rough agreement with the experimental 
results [9]. Note that the finite resolution of the imaging 
process reduces the number of detected solitons [18]. 
\par 
Finally, we observe that Eq. (24) 
has been obtained by using NLSE. NPSE gives that 
same results of NLSE in the quasi-1D limit 
(of experiment [9]) but predicts the formation 
a large number of solitons also close to the critical 
point $g n = 1$. The consequences of this prediction 
will be discussed elsewhere. 

\section*{Conclusions}

In this paper we have shown that the solitary-waves studied in [8] 
belong to the stable branch $-1<\omega<-1/2$ we have here 
analytically found by using the Vakhitov-Kolokolov criterion. 
In this branch, growing $\omega$ the number of particles 
$N_s$ increases and for a fixed $N_s$ the solitary-wave solution 
is fully determined. 
Since the study of the 3D model is usually time consuming, it is nice to 
have a simple 1D model to understand the dynamics of a Bose-Einstein 
condensate. NPSE has the same numerical complexity of other 
one-dimensional nonlinear Schr\"odinger equations but we have 
shown that it is not well approximated by the cubic or 
cubic-quintic Schr\"odinger equations. 
NPSE reflects correctly the 3D dynamics of Bose condensed 
solitary-waves and it predicts with considerable 
accuracy the collapse of the condensate. 
Moreover, we have shown the NPSE explains 
remarkably well the formation of multiple bright solitons 
by a sudden change of the sign of the 
nonlinear strength from positive to negative, as 
a consequence the modulational instability of the 
time-dependent wave function.

\newpage

\begin{figure}
\centerline{\psfig{file=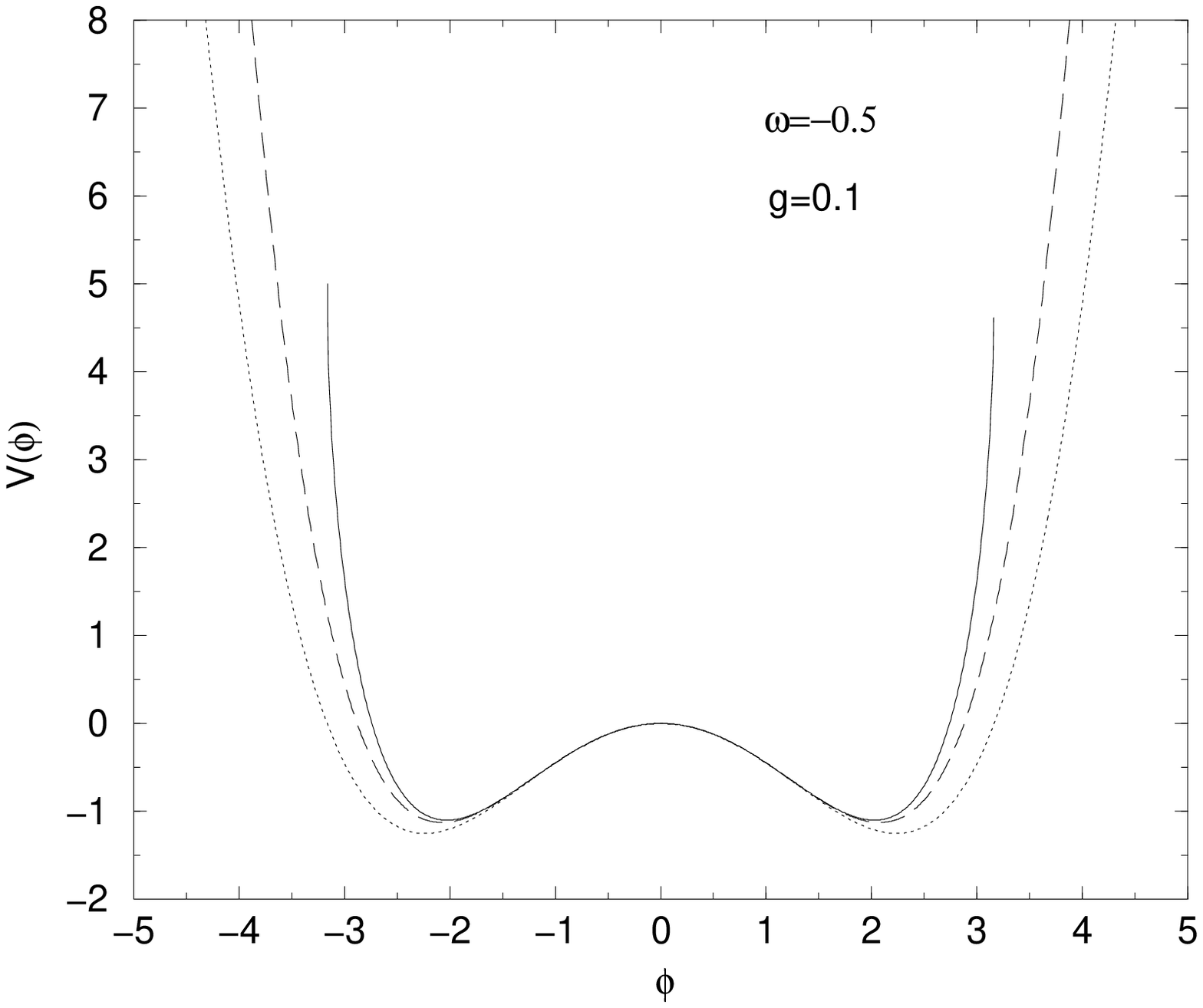,height=5.4in}}
\caption{Effective potential energy $V(\phi)$ 
of the NPSE. Exact: full line; cubic approximation: 
dotted line, cubic-quintic approximation: dashed line. 
$\omega$ is the phase frequency and $g$ is the nonlinear 
strength.} 
\end{figure}

\newpage 

\begin{figure}
\centerline{\psfig{file=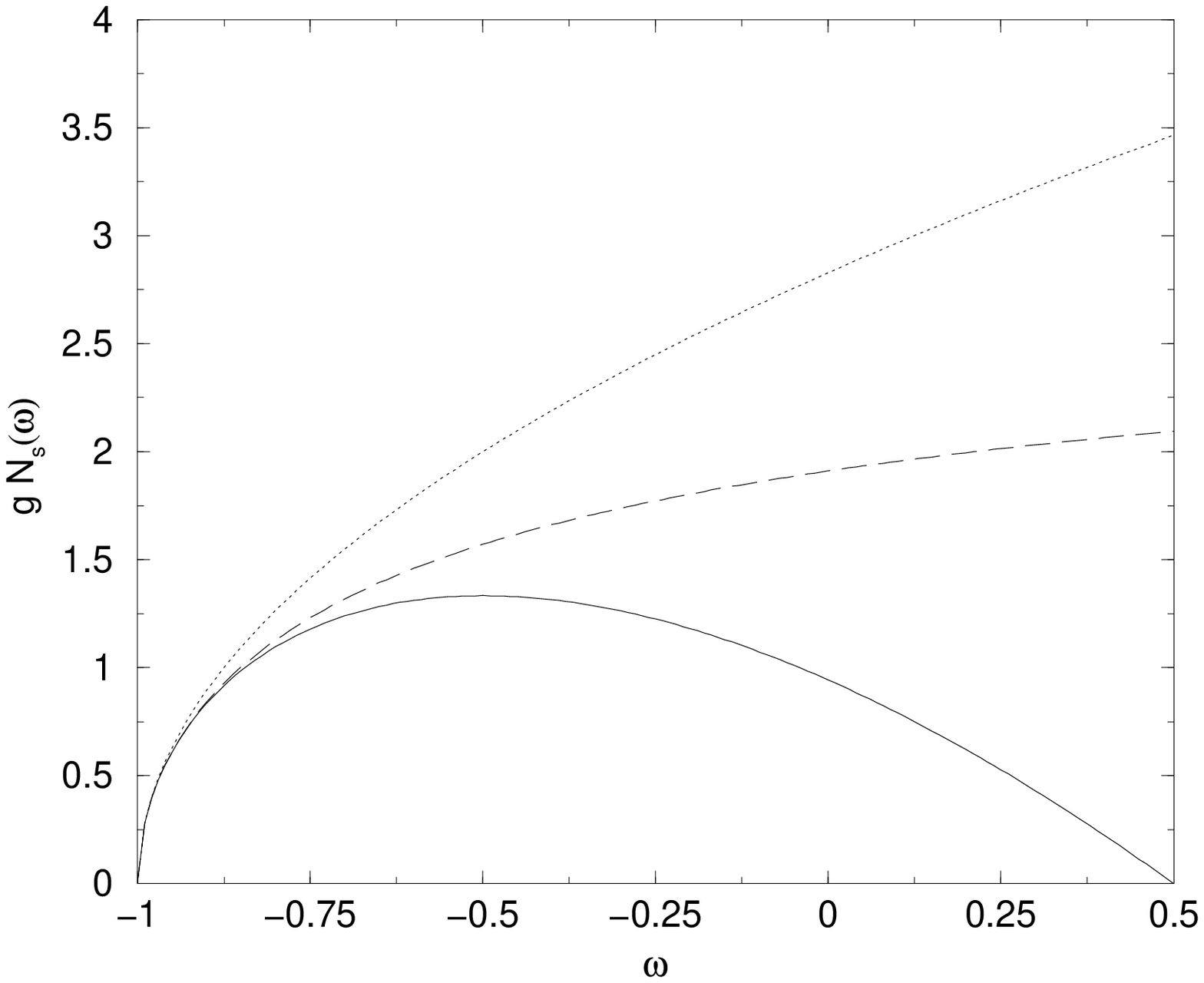,height=5.4in}}
\caption{Number of particles $N_s(\omega )$ as a function 
of the phase frequency $\omega$ for the NPSE. 
Exact: full line; cubic approximation: dotted line; 
cubic-quintic approximation: dashed line. 
$g$ is the nonlinear strength.} 
\end{figure}


\begin{thebibliography}{99}

\bibitem{rf:1} L. Salasnich, Laser Phys. {\bf 12}, 198 (2002). 

\bibitem{rf:2} L. Salasnich, A. Parola, and L. Reatto,  
Phys. Rev. A {\bf 65}, 043614 (2002). 

\bibitem{rf:3} L. Salasnich, A. Parola, and L. Reatto, 
J. Phys. B {\bf 35}, 3205 (2002); 
L. Salasnich, Laser Phys. {\bf 13}, 543 (2003).   

\bibitem{rf:4} A.J. Leggett, Rev. Mod. Phys. 
{\bf 73}, 307 (2002). 

\bibitem{rf:5} P.W. Courteille, V.S. Bagnato, and 
Y.I. Yukalov, Laser Phys. {\bf 11}, 659 (2001). 

\bibitem{rf:6} E.P. Gross, Nuovo Cimento {\bf 20}, 454 (1961). 

\bibitem{rf:7} L.P. Pitaevskii, Zh. Eksp. Teor. Fiz. {\bf 40}, 
646 (1961) [English Transl. Sov. Phys. JETP {\bf 13}, 451 (1961)].

\bibitem{rf:8} L. Salasnich, A. Parola, and L. Reatto, 
Phys. Rev. A {\bf 66} (2002). 

\bibitem{rf:9} K.E. Strecker {\it et al.}, 
Nature {\bf 417}, 150 (2002). 

\bibitem{rf:10} L. Khaykovich {\it et al.}, Science {\bf 296}, 
1290 (2002). 

\bibitem{rf:11} M.G. Vakhitov and A.A. Kolokolov, Izv. Vyssh. Uch. Zav. 
radiofizika {\bf 16}, 1020 (1973) 
[English Transl.Radiophys. Quantum Electron {\bf 16}, 783 (1973)]. 

\bibitem{rf:12} M.I. Weinstein, Comments Pure Appl. Math. {\bf 39}, 
51 (1986). 

\bibitem{rf:13} D.E. Pelinovsky, V.V. Afanasjev, and Y.S. Kivshar, 
Phys. Rev. E {\bf 53}, 1940 (1996). 

\bibitem{rf:14} A. Gammal, L. Tomio, T. Federico, 
Phys. Rev. A {\bf 66}, 043619 (2002). 

\bibitem{rf:15} U. Al Khawaja, H.T.C. Stoof, R.G. Hulet, K.E. Strecker, 
and G.B. Partridge, Phys. Rev. Lett. {\bf 89}, 200404 (2002). 

\bibitem{rf:16} T. Taniuti and H. Washimi, Phys. Rev. Lett. {\bf 21}, 
209 (1968); A. Hasegawa, Phys. Rev. Lett. {\bf 24}, 1165 (1970). 

\bibitem{rf:17} L.A. Ostrovskii, Sov. Phys. JEPT {\bf 24}, 797 (1969); 
K. Tai, A. Hasegawa, and A. Tomita, Phys. Rev. Lett. {\bf 56}, 135 (1986). 

\bibitem{rf:18} L. Salasnich, A. Parola, and L. Reatto, 
``Modulational instability and complex dynamics of 
confined matter-wave solitons'', submitted for publication. 

\end{thebibliography}
\end{document}